\documentclass[a4paper]{article}

\usepackage{icrc2013}

\title{The $\mathbf{\langle \ln A \rangle}$ study in the primary energy range $\mathbf{10^{16}}$~eV~-~$\mathbf{10^{17}}$~eV with the Muon Tracking Detector in the KASCADE-Grande experiment}

\shorttitle{$\langle \ln A \rangle$ from MTD in KASCADE-Grande}

\authors{
P.~{\L}uczak$^{1}$,
W.D.~Apel$^{2}$,
J.C.~Arteaga-Vel\'azquez$^{3}$,
K.~Bekk$^{2}$,
M.~Bertaina$^{4}$,
J.~Bl\"umer$^{2,5}$,
H.~Bozdog$^{2}$,
I.M.~Brancus$^{6}$,
E.~Cantoni$^{4,7,a}$,
A.~Chiavassa$^{4}$,
F.~Cossavella$^{5,b}$,
C.~Curcio$^{4}$,
K.~Daumiller$^{2}$,
V.~de Souza$^{8}$,
F.~Di~Pierro$^{4}$,
P.~Doll$^{2}$,
R.~Engel$^{2}$,
J.~Engler$^{2}$,
B.~Fuchs$^{5}$,
D.~Fuhrmann$^{9,c}$,
H.J.~Gils$^{2}$,
R.~Glasstetter$^{9}$,
C.~Grupen$^{10}$,
A.~Haungs$^{2}$,
D.~Heck$^{2}$,
J.R.~H\"orandel$^{11}$,
D.~Huber$^{5}$,
T.~Huege$^{2}$,
K.-H.~Kampert$^{9}$,
D.~Kang$^{5}$, 
H.O.~Klages$^{2}$,
K.~Link$^{5}$, 
M.~Ludwig$^{5}$,
H.J.~Mathes$^{2}$,
H.J.~Mayer$^{2}$,
M.~Melissas$^{5}$,
J.~Milke$^{2}$,
B.~Mitrica$^{6}$,
C.~Morello$^{7}$,
J.~Oehlschl\"ager$^{2}$,
S.~Ostapchenko$^{2,d}$,
N.~Palmieri$^{5}$,
M.~Petcu$^{6}$,
T.~Pierog$^{2}$,
H.~Rebel$^{2}$,
M.~Roth$^{2}$,
H.~Schieler$^{2}$,
S.~Schoo$^{2}$,
F.G.~Schr\"oder$^{2}$,
O.~Sima$^{12}$,
G.~Toma$^{6}$,
G.C.~Trinchero$^{7}$,
H.~Ulrich$^{2}$,
A.~Weindl$^{2}$,
J.~Wochele$^{2}$,
J.~Zabierowski$^{1}$ \\
KASCADE-Grande Collaboration
}

\afiliations{
$^1$ National Centre for Nuclear Research, Department of Cosmic Ray Physics, {\L}{\'o}d{\'z}, Poland\\
$^2$ Institut f\"ur Kernphysik, KIT - Karlsruher Institut f\"ur Technologie, Germany\\
$^3$ Universidad Michoacana, Instituto de F\'{\i}sica y Matem\'aticas, Morelia, Mexico\\
$^4$ Dipartimento di Fisica, Universit\`a degli Studi di Torino, Italy\\
$^5$ Institut f\"ur Experimentelle Kernphysik, KIT - Karlsruher Institut f\"ur Technologie, Germany\\
$^6$ Horia Hulubei National Institute of Physics and Nuclear Engineering, Bucharest, Romania\\
$^7$ Osservatorio Astrofisico di Torino, INAF Torino, Italy\\
$^8$ Universidade S$\tilde{a}$o Paulo, Instituto de F\'{\i}sica de S\~ao Carlos, Brasil\\
$^9$ Fachbereich Physik, Universit\"at Wuppertal, Germany\\
$^{10}$ Department of Physics, Siegen University, Germany\\
$^{11}$ Dept. of Astrophysics, Radboud University Nijmegen, The Netherlands\\
$^{12}$ Department of Physics, University of Bucharest, Bucharest, Romania\\
\scriptsize{
$^{a}$ now at: Istituto Nazionale di Ricerca Metrologia, INRIM, Torino; \\
$^{b}$ now at: Max-Planck-Institut Physik, M\"unchen, Germany; \\
$^{c}$ now at: University of Duisburg-Essen, Duisburg, Germany;  \\
$^{d}$ now at: University of Trondheim, Norway.
}
}

\email{luczak@zpk.u.lodz.pl} 

\abstract{The KASCADE-Grande Muon Tracking Detector enables with high accuracy the measurement of directions of EAS muons with energy above 0.8~GeV and up to 700~m distance from the shower centre. Reconstructed muon tracks are used to investigate muon pseudorapidity ($\eta$) distributions. These distributions are nearly identical to the pseudorapidity distributions of their parent mesons produced in hadronic interactions. Comparison of the $\eta$ distributions from measured and simulated showers can be used to test the quality of the high energy hadronic interaction models. In this context a comparison of the QGSJet-II-2 and QGSJet-II-4 model will be shown. The pseudorapidity distributions reflect the longitudinal development of EAS and, as such, are sensitive to the mass of the cosmic rays primary particles. With various parameters of the $\eta$ distribution, obtained from the MTD data, it is possible to calculate the mean logarithmic mass of CRs. The results of the $\langle \ln A \rangle$ analysis in the primary energy range $10^{16}$~eV~-~$10^{17}$~eV with the $1^{st}$~quartile (Q1) of $\eta$ distribution will be presented. \\}

\vspace{-5cm}

\begin{document}
\maketitle

%Begin a section.

\section{Introduction}

\begin{figure*}[th]
\begin{center}
\centering
\includegraphics*[width=\textwidth]{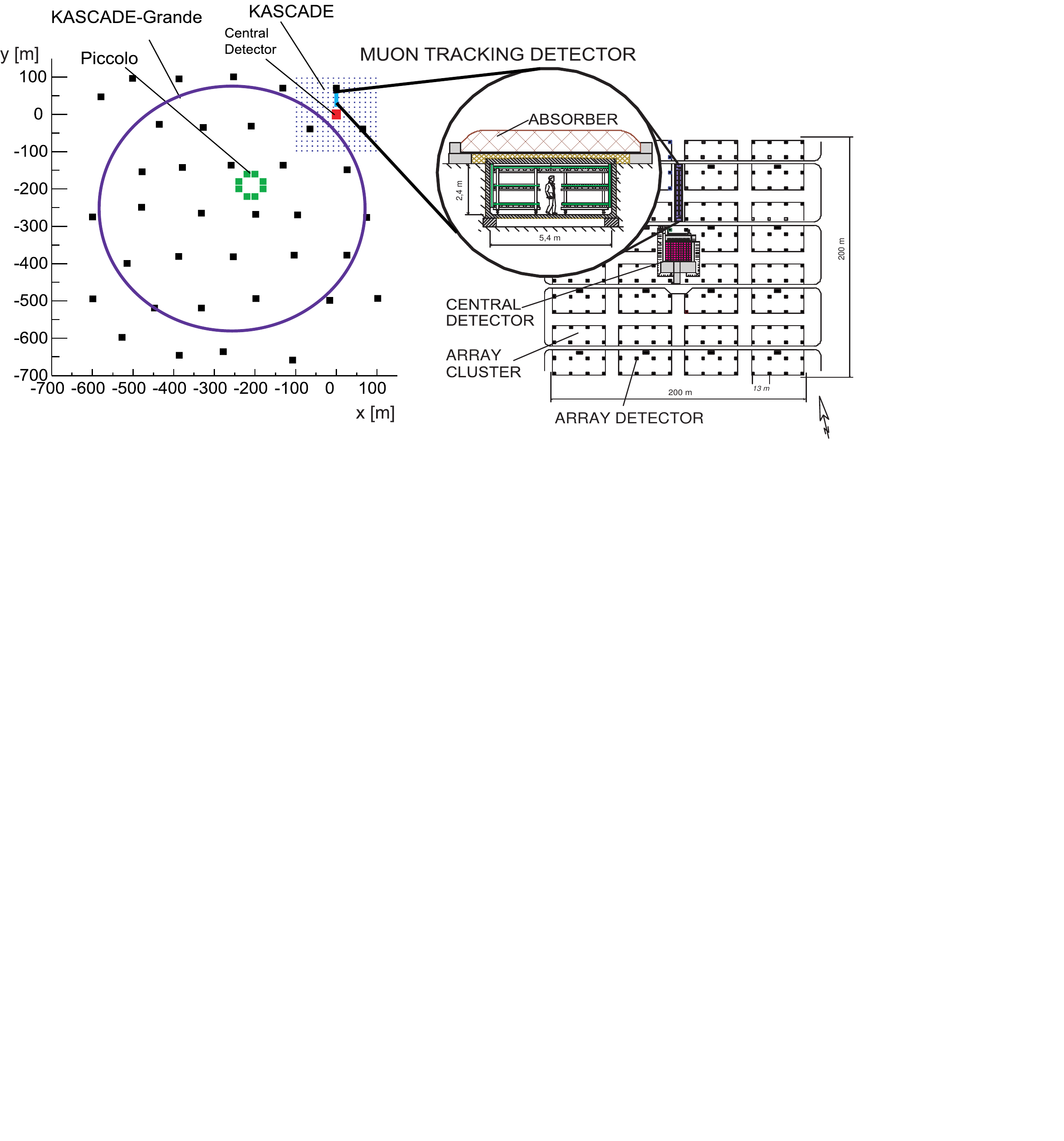}
\vspace{-12cm}
\caption{Layout of the KASCADE-Grande experiment distributed over the KIT - Campus North area.
KASCADE is situated in the North-East corner of the Campus; note the position of the Muon Tracking Detector.}
\label{fig1}
\end{center}
\end{figure*}

\indent The Muon Tracking Detector (MTD) \cite{MTD} is one of the detector components in the KASCADE-Grande EAS experiment \cite{KASG} (see figure~\ref{fig1}), operated at the Karlsruhe Institute of Technology (KIT) - Campus North, in Germany, by an international collaboration. \\
\indent The MTD is a group of 16 muon telescopes made out of streamer tube gas detectors grouped in four modules (3 horizontal, 1 vertical). Each module is 2~meters wide and 4~meters long, the total detection area for vertical particles is 128~m$^{2}$. The telescopes are located in a concrete tunnel, under the ground level, covered by the concrete-soil-iron absorbent which absorbs a large fraction of accompanying low-energy particles, thus enhancing the identification of muons with an energy exceeding 0.8~GeV. The MTD measures directions of muon tracks in extensive air showers with \mbox{excellent} angular resolution of $\approx$~ $0.35^{\circ}$.\\
\indent These directional data enables to investigate the longitudinal development of the muonic component in air showers which is a signature of the development of the hadronic EAS core, being in turn dependent on the mass of the primary cosmic ray particle initiating a shower. Such studies can be done either by the determination of a mean muon production height \cite{Hmu} or by using the mean pseudorapidity ($\eta$) of EAS muons, expressed in terms of their tangential ($\tau$) and radial ($\rho$) angles, quantities reconstructed from muon tracks obtained with the MTD \cite{eta}, \cite{pylos}. \\
\indent In this work we compare the $\eta$ distributions for two high energy hadronic interaction models and investigate to what extent one can use the value of the first quartile of the muon pseudorapidity distribution for the determination of the mean logarithm of mass of cosmic rays.

\section{Comparison of the $\eta$ distributions for QGSJet-II-2 and QGSJet-II-4}

 \begin{figure}[t]
  \centering
  \includegraphics[trim=2.2cm 7cm 3.7cm 7cm, clip=true,width=0.4\textwidth]{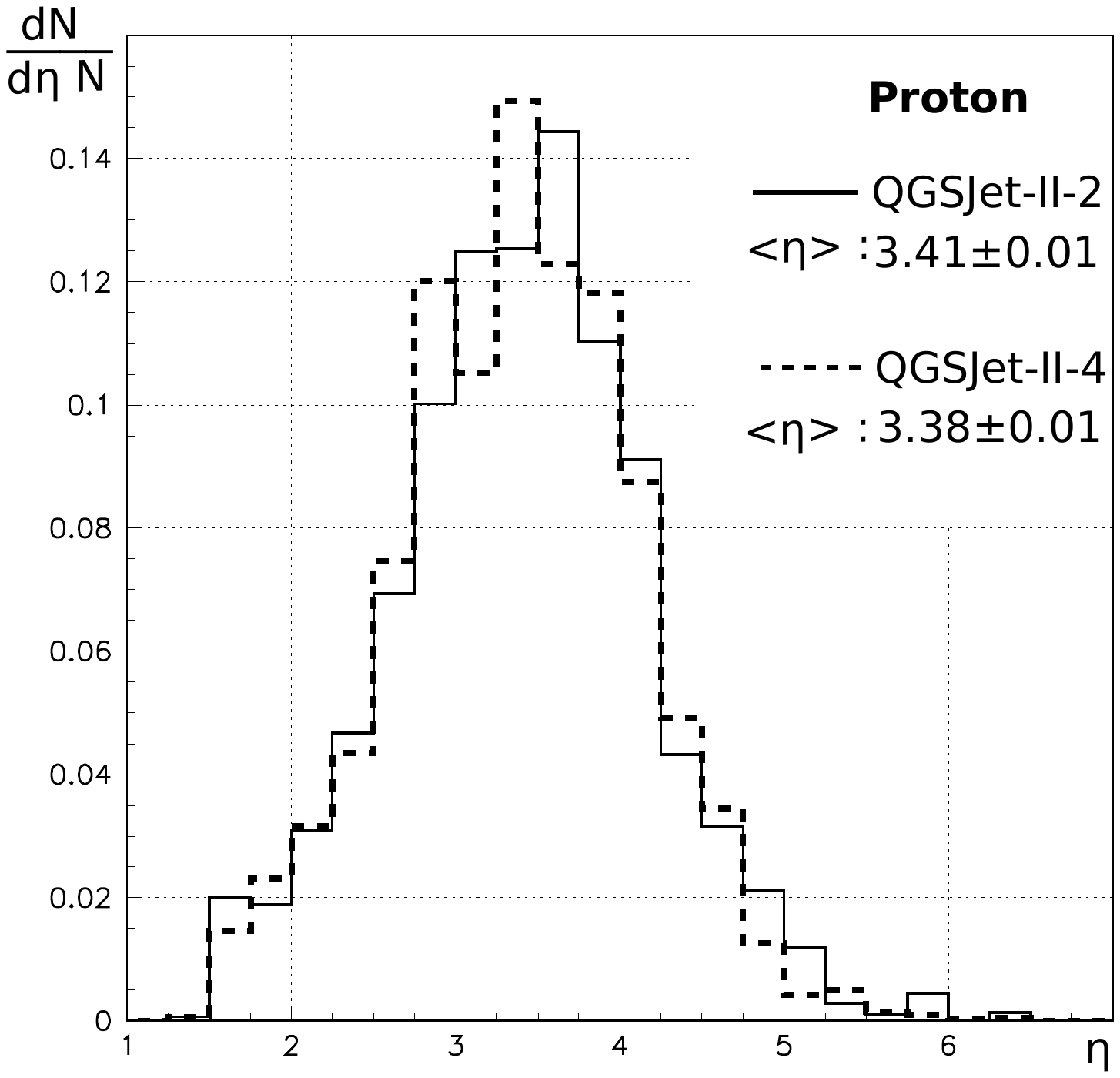}
  \caption{Comparison between pseudorapidity distributions for proton initiated air showers simulated with QGSJet-II-2 (solid line) and QGSJet-II-4 (dashed line)~-~see text.}\label{fig3}
 \end{figure}
 \begin{figure}[t]
  \centering
  \includegraphics[trim=2.2cm 7cm 3.7cm 7cm, clip=true,width=0.4\textwidth]{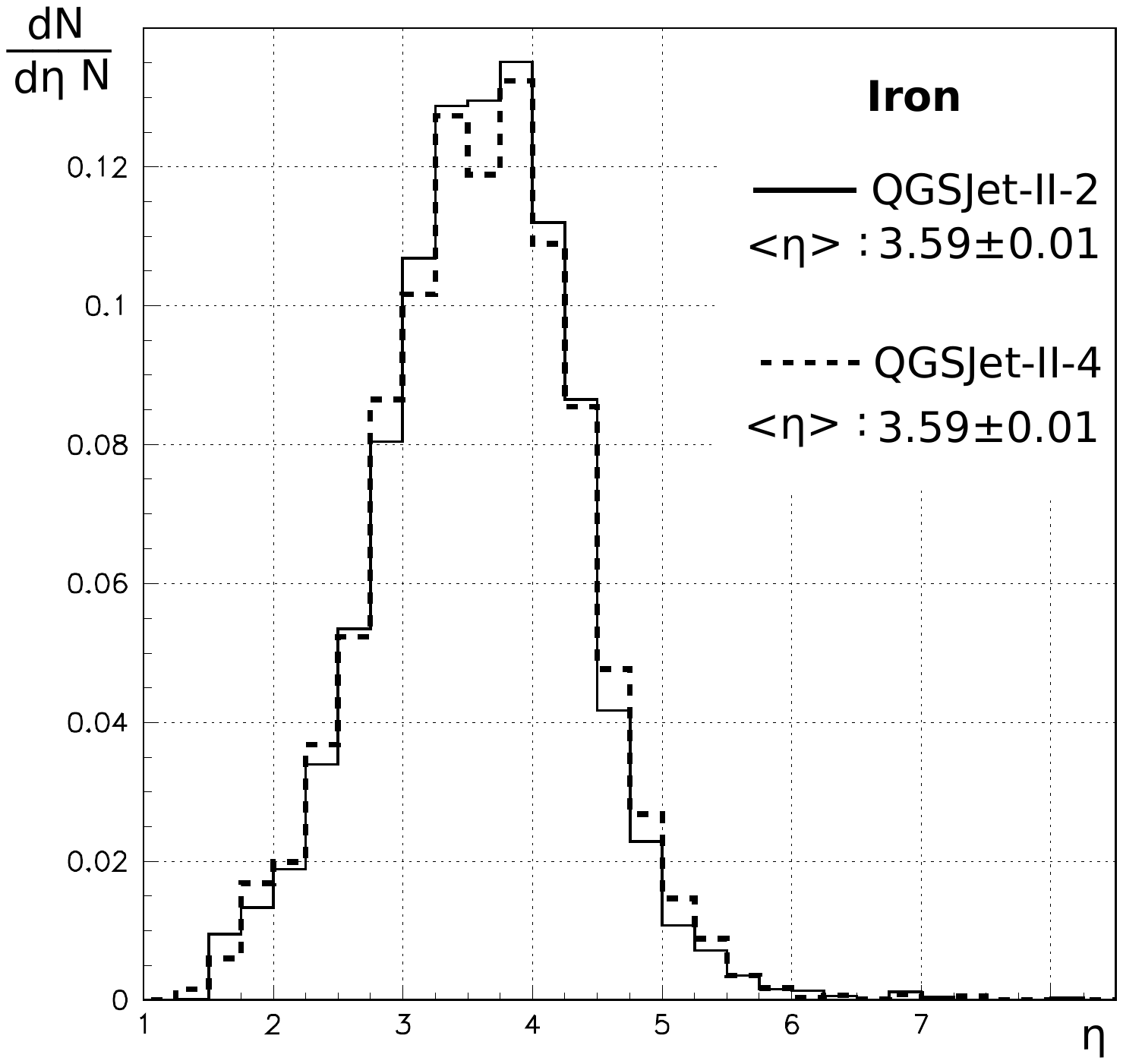}
  \caption{Comparison between pseudorapidity distributions for iron initiated air showers simulated with QGSJet-II-2 (solid line) and QGSJet-II-4 (dashed line)~-~see text.}\label{fig4}
 \end{figure}

  \indent In this analysis EAS initiated by the proton and iron primary CR particles were simulated with the \mbox{CORSIKA}\cite{heck} code. The showers were simulated with two combinations of hadronic interaction models \mbox{QGSJet-II-2}\footnote{Ref. \cite{qgs}}+FLUKA2002.4\footnote{Ref. \cite{fluk}} and \mbox{QGSJet-II-4}\footnote{Ref. \cite{qgs4}}+FLUKA2012.2.14. The output of the CORSIKA was analysed with \mbox{CRES}\footnote{Simulation of the detector response.} and \mbox{KRETA}\footnote{Reconstruction and analysis of the shower parameters.} codes in the same way as the measured data are analysed. The comparison between both simulation sets is done using similar number of showers simulated and analysed under the same initial conditions.\\
  \indent Figures \ref{fig3} and \ref{fig4} show the comparison between $\eta$ distribution for showers simulated with the \mbox{QGSJet-II-2}+\mbox{FLUKA} (solid line) and the \mbox{QGSJet-II-4}+\mbox{FLUKA} (dashed line) for proton and iron initiated showers. Both distributions are obtained for -0.5$^{\circ}<\rho<23^{\circ}$ and 0$^{\circ}<|\tau|<23^{\circ}$ in the muon-to-shower-axis distance range 250-370~meters, for showers with zenith angle up to 18$^{\circ}$ and in the energy range $\lg(E_{0}^{rec}[GeV])>7.0$. Due to the low statistics, there are no statistically significant differences between the $\eta$ distributions for both models which was checked by performing a Kolmogorov-Smirnov test for the two distributions. The shapes of the distributions are similar for both primary particles, but one can notice that the mean $\eta$ of the proton distribution is lower for the QGSJet-II-4 model. This shift in proton distribution should be confirmed with more statistics of showers. If it remains, it will improve the se\-pa\-ra\-tion between proton, data and iron distributions, leading to more reliable description of the measured data and higher values of the $\langle \ln A \rangle$ calculated from the parameters of the $\eta$ distributions.\\

\section{The mass sensitivity of the EAS muon pseudorapidity}
\indent In the KASCADE-Grande experiment muons can be re\-gi\-ste\-red up to 700~meters from the shower core. In case of the \mbox{presented} analysis, the muon-to-shower-axis (R$_{\mu}$) distance ranges (Table~\ref{tab2}) were limited to the distances where the mass composition of the detected showers is constant (taking into account the $\lg(N_{\mu})$/$\lg(N_{e})$ ratio) in each energy range, not being affected by experimental inefficiencies. In this analysis EAS initiated by the proton and iron primary CR particles were simulated with the CORSIKA code (\mbox{QGSJet-II-2} as high energy hadronic (HE) interaction model and \mbox{FLUKA2002.4} as low energy (LE) hadronic interaction model). The QGSJet model was used to simulate hadronic interactions of particles with energy above 200~GeV while the FLUKA model was used to simulate ones below this energy. In the measured and simulated data only showers with zenith angle up to 18$^{\circ}$ are analysed.\\
\begin{table}[!b]
\caption{R$_{\mu}$ distance ranges for each analysed energy range.}
\begin{center}
\begin{tabular}{ccc}
$\lg(E_{0}^{rec}[GeV])$&$\langle E^{rec}_{0} \rangle$	&R$_{\mu}$ [m] \\
&10$^{7}$ GeV			&\\
 \hline
7.0-7.3&1.34$\pm$0.01&250-370  \\
$>$7.0 &2.07$\pm$0.01&250-370  \\
7.3-7.6&2.69$\pm$0.01&280-400 \\
7.6-7.9&5.34$\pm$0.02&280-430\\
$>$7.6 &7.15$\pm$0.07&280-430  \\
\hline
\end{tabular}
\label{tab2}
\end{center}
\end{table}
\indent It was shown in \cite{phd} that the pseudorapidity of EAS muons is a parameter sensitive to the mass of the primary cosmic ray particles and can be used to calculate the $\langle \ln A \rangle$ of CRs. 
It was observed that the $\eta$ distributions from the measured data are bracketed by the distributions from simulated showers and that $\langle \ln A \rangle$ and $\langle \eta \rangle$ are linearly dependent. The main conclusion of the analysis is that $\langle \ln A \rangle$ is increasing with the energy, but its values are lower than expected in the energy range $10^{16}$~eV-$10^{17}$~eV. The reason for this are the distortions of the $\eta$ distributions in simulations caused by the large number of high $\eta$ muons that are shifting the $\langle \eta \rangle$ of the distributions towards higher values. As a result of this shift, the distributions from the measurement are close to these from proton initiated showers. This behaviour is caused by the QGSJet model which creates too many high $\eta$ muons that reach the observation level.\\
 \begin{figure}[!b]
  \centering
  \includegraphics[trim=2.2cm 10cm 3.7cm 10.5cm, clip=true,width=0.4\textwidth]{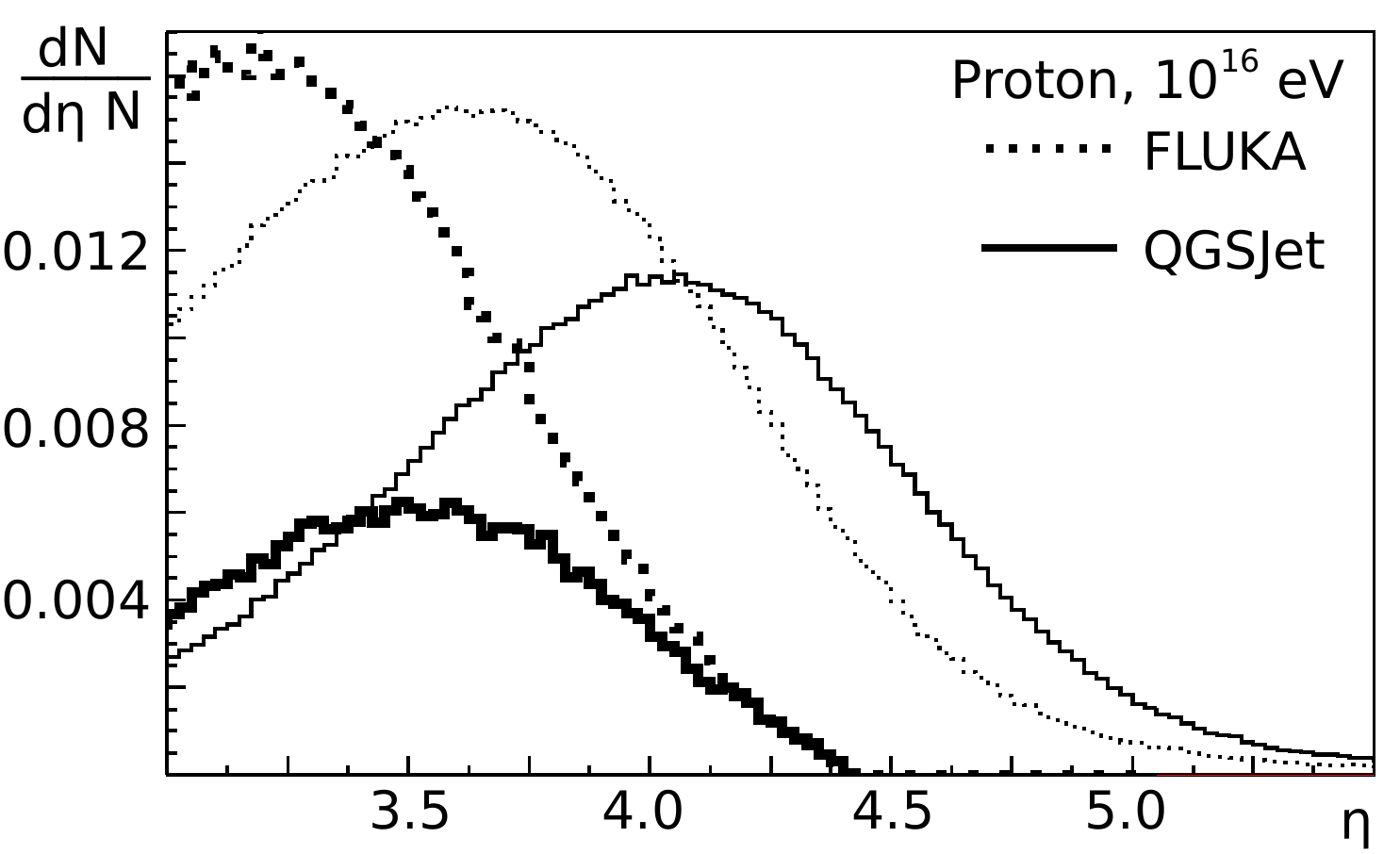}
  \caption{Comparison between $\eta$ distributions from HE (solid lines, QGSJet) and LE (dotted lines, FLUKA) muon sample before (thin) and after (bold)  the angular cuts~-~see text.}
  \label{fig2}
 \end{figure}
\indent This conclusion was confirmed by the investigation of the muon production height (h$_{\mu}$)\cite{Hmu}. In this analysis, h$_{\mu}$ of muons, from proton and iron initiated showers, were compared with muons from detected showers divided into these from light and heavy primary CR particles. It occurred that the measured values of the h$_{\mu}$ above 4~km, where the interactions are described by the HE interaction model are not well described by the simulations. This is not the case for h$_{\mu}<$4~km, where interactions are described by the LE interaction model, and where the measurements are compatible with the simulations. \\
\indent In the simulations it is possible to distinguish between muons that stem from hadrons originating from HE or LE hadronic interaction models. Dividing the muon sample into those originating from grandparent hadrons with energy above or below 200~GeV (HE and LE muon sample) we can analyse contributions of these two groups to the combined $\eta$ distribution which can be obtained with the MTD data. In the R$_{\mu}$ distance range that is valid in case of the MTD analysis (250-400~meters), there are about 70\% muons from the LE sample and 30\% from the HE sample. Most of the muons from the latter contribute to the $\eta>4$ of the distribution, creating a long tail of high $\eta$ muons. In this pseudorapidity range the number of muons from this HE sample is larger than from the LE sample (with ratio  about 60\% to 40\%, respectively). \\

\indent The differences in the LE and HE muon sample contributions suggest that calculating the $\langle \ln A \rangle$ with the $1^{st}$ quartile of the distribution, where LE interaction model dominates, will remove the influence of the HE interaction model. However, the value of the quartile is affected by the tail from the HE muon sample and the analysis is biased. To eliminate this bias  the radial and tangential angle cuts were applied. The effect of such angular cuts is depicted in figure~\ref{fig2}. Here the $\eta$ distributions from the QGSJet and FLUKA muon sample (solid line and dotted line, respectively) are compared before and after the angular cuts (thin and bold solid lines and dotted lines, respectively). However, in the ex\-pe\-ri\-ment conditions, it is necessary to take into account the statistics of available simulated and measured data. That is why it was not possible to eliminate completely the tails from the HE muon sample.

\indent The angular cuts which provide as \mbox{small} as possible decrease in muon statistics and suppress tails from HE sample without significant distortion of the shape of the pseudorapidity distribution are: 0.75$^{\circ}<\rho<17^{\circ}$ and 0.2$^{\circ}<|\tau|<17^{\circ}$.\\ 

\section{Results and conclusions}

\indent Comparison between $\eta$ distributions from H and Fe showers simulated  with QGSJet-II-2 and QGSJet-II-4 model revealed that with the currently available statistics it is too early to draw conclusions about the quality of the new version of the QGSJet model. However, the shift in the $\langle \eta \rangle$ of the proton distribution indicates that with larger number of simulated showers we can expect that a change in separation between proton and iron distributions, will lead to the better description of the measured data by the simulations.\\

\begin{figure}[!h]
  \centering

\includegraphics[trim=4cm 12cm 5cm 10cm, clip=true,width=0.5\textwidth]{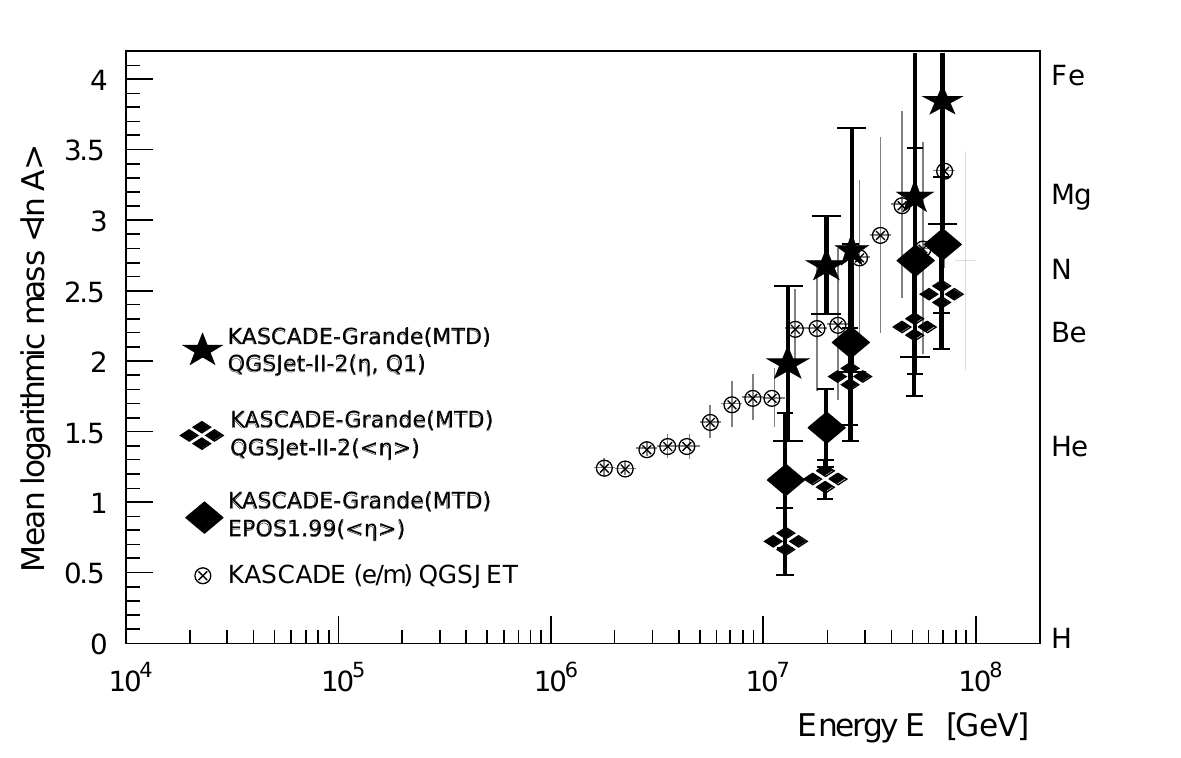}
%\vspace{-1cm}
\caption{Results of the $\langle \ln A \rangle$ values obtained with the mean and the first quartile(preliminary) from the muon pseudorapidity distributions, shown together with the values derived from the measurements of electrons and muons in the air showers obtained with the KASCADE detectors\cite{antoni2005}.}
\label{fig5}
 \end{figure}

\indent Combining angular cuts and the $1^{st}$ quartile analysis, the calculated values of the $\langle \ln A \rangle$ rise with the energy at a similar rate as the ones from electron/muon detectors in KASCADE, and are significantly higher than in the previous analysis \cite{phd}. These values are similar to the $\langle \ln A \rangle$ results obtained by KASCADE with the e/m ratio analysis using QGSJet model \cite{antoni2005}.\\ 
\indent The results of the calculations are presented in Table~\ref{tablelnajvbqfx}, and compared with the results from the previous analysis where the $\langle \eta \rangle$ values of the distributions were used. \mbox{Figure~\ref{fig5}} shows the results from the table compared with the results from electron/muon detectors in the KASCADE experiment.
\begin{table}[!h]
  \caption{Comparison between $\langle \ln A \rangle$ values calculated with the $1^{st}$ quartile and the mean pseudorapidity from the $\eta$  distributions.}
  \begin{center}
  \begin{tabular}{c			 c|				c|			c}	
		  $\lg(E^{rec}_{0}[GeV])$&$\langle E^{rec}_{0} \rangle$	&\multicolumn{2}{c}{{$\langle \ln A \rangle$, QGSJet-II-2}}\\
					 &10$^{7}$ GeV			&$\eta,~Q1$	&$\langle \eta \rangle$\\
  \hline
  7.0 - 7.3	&1.34$\pm$0.01&1.98$\pm$0.55&0.71$\pm$0.24\\
  $>$7.0	&2.07$\pm$0.01&2.68$\pm$0.36&1.15$\pm$0.14\\
  7.3 - 7.6	&2.69$\pm$0.01&2.79$\pm$0.87&1.87$\pm$0.34\\
  7.6 - 7.9	&5.34$\pm$0.02&3.17$\pm$1.12&2.13$\pm$0.48\\
  $>$7.6	&7.15$\pm$0.07&3.86$\pm$0.81&2.45$\pm$0.39\\
  \hline
  \end{tabular}
  \label{tablelnajvbqfx}
  \end{center}
\end{table}

\vspace*{0.99cm}
\footnotesize{{\bf Acknowledgment:}{The authors would like to thank the members of the
engineering and technical staff of the KASCADE-Grande
collaboration, who contributed to the success of the ex\-pe\-ri\-ment.
The KASCADE-Grande ex\-pe\-ri\-ment is supported
in Germany by the BMBF and by the Helmholtz Alliance
for Astroparticle Physics - HAP funded by the Initiative
and Networking Fund of the Helmholtz Association, 
by the MIUR and INAF of Italy,
the Polish Ministry of Science and Higher Education,
and the Romanian Authority for Scientific Research UEFISCDI 
(grants PNII-IDEI 271/2011 and RU-PD 17/2011).
}}

\end{document}